# Metallo-dielectric hybrid antennas for ultrastrong enhancement of spontaneous emission

Xue-Wen Chen[1], Mario Agio[2] and Vahid Sandoghdar[1]

[1] Max Planck Institute for the Science of Light, D-91058 Erlangen, Germany

[2] National Institute of Optics (CNR-INO) and European Laboratory for Nonlinear Spectroscopy (LENS)

via Nello Carrara 1, 50019 Sesto Fiorentino (FI), Italy

## ABSTRACT

We devise new optical antennas that reduce the excited-state radiative lifetimes of emitters to the order of 100 femtoseconds while maintaining quantum efficiencies of about 80% at a broadband operation. Here, we combine metallic nanoparticles with planar dielectric structures and exploit design strategies from plasmonic nanoantennas and concepts from Cavity Quantum Electrodynamics to maximize the local density of states and minimize the nonradiative losses incurred by the metallic constituents. The proposed metallo-dielectric hybrid antennas promise important impact on various fundamental and applied research fields, including photophysics, ultrafast plasmonics, bright single photon sources and Raman spectroscopy.

Modification of spontaneous emission has been a topic of theoretical and experimental studies over the past four decades and has given birth to the field of Cavity Quantum Electrodynamics (CQED) [1, 2]. The central idea in this research area is to manipulate the local photonic density of states (LDOS) at the position of an emitter by placing material boundaries in its surrounding. To achieve this, the early efforts considered flat interfaces [3, 4] and Fabry-Perot resonators [1, 2], while other works have examined structured dielectric material [5] and nanoparticles [6-8]. In particular, recent reports have shown that plasmonic nanoantennas can enhance emission and reception of radiation in analogy with antenna concepts from radio engineering [9-13]. In this Letter, we theoretically show that combinations of simple plasmonic nanoparticles and planar dielectric interfaces can reduce the electronic excited-state lifetimes of atoms from the typical values of the order of nanoseconds to the range of 100 femtoseconds.

To set the ground, we consider a single quantum emitter close to a subwavelength metallic sphere (see Fig. 1a), serving as a simple metallic nanoantenna (MNA) [10]. If the transition dipole moment of the emitter is oriented radially, it polarizes the MNA, giving rise to a large total dipole moment in the coupled system. To maximize this coupling, and thus the overall radiative rate ($\gamma_{\text{rad}}$), it is desirable to decrease the MNA-emitter separation. However, for very small distances, the electric field in the MNA strongly deviates from a homogeneous distribution. This reduces the dipolar polarization of the MNA and enhances the nonradiative decay rate ($\gamma_{\text{nr}}$) caused by dissipation in the metal. In a previous work, we pointed out that one could alleviate such fluorescence quenching by shifting the plasmon resonance to the near infrared, where metals absorb considerably less than in the visible spectrum [14]. We also pointed out that an important antenna design criterion should be a large dipolar polarizability of the MNA. In our current investigation, we extend those concepts by introducing planar dielectric interfaces to amplify the strength of the electric field at the position of the MNA, improve its homogeneity within the MNA, and increase the density of states for its radiation even in the visible regime.

The spontaneous emission rate of an atom embedded in a homogeneous bulk dielectric of refractive index *n* scales as $\gamma_{\text{bulk}} = n\gamma_{\text{vac}}$ [15, 16], where $\gamma_{\text{vac}}$ denotes the spontaneous emission rate of the emitter in vacuum. If an interface is introduced close to the emitter, its decay rate ($\gamma_{\text{int}}$) becomes dependent on its orientation and position [4]. For example, $\gamma_{\text{int}}$ of an atom embedded in a dielectric with $n$=2.5 and at a depth of $d$=4 nm beneath its interface with air is reduced to $0.18\gamma_{\text{vac}}$ if the emission dipole moment is normal to the interface, as shown in Fig. 1b. Conversely, the decay rate is enhanced if the emitter sits in the low-index part [17]. The resulting effect can be attributed to the modification of the LDOS caused by the boundary conditions at the dielectric interface. Although this arrangement does not involve any cavity, it has been commonly discussed in the context of CQED [2, 18]. In light of the new developments in nano-optics [9-13] we find it more appropriate to think of the interface as a dielectric planar antenna (DPA), which mediates between the near field of the emitter and its far-field radiation. Indeed, the strongly modified radiation pattern of an emitter in this geometry [19-21] is quite reminiscent of an antenna effect.

We now combine two simple MNA and DPA as shown in Fig. 1c, where a 100 nm gold nanosphere is placed in contact with an infinite dielectric substrate containing a vertically-oriented emitter embedded at distance $d$=4 nm below the interface. Our computation of the spontaneous emission rate and radiation efficiency throughout this work is based on accurate and efficient body-of-revolution finite-difference time domain (BOR-FDTD) calculations [22], where we take advantage of the rotation symmetry of the system to implement very fine grids (0.1 nm). Figures 1d and 1e depict the time-averaged z-component of the electric field intensity generated by a vertical electric dipole placed in a media with $n$=2.5 and $n$=1, respectively. A remarkable feature of these results is that the field intensity in Fig. 1d is more than one order of magnitude larger than its counterpart in Fig. 1e, indicating much higher emitted power and thus faster emission.

Figures 2a and 2b display the spectral dependencies of $\gamma_{\text{rad}}$ and the quantum efficiency $\eta = \frac{\gamma_{\text{rad}}}{\gamma_{\text{rad}}+\gamma_{\text{nr}}}$, respectively, for an emitter coupled to hybrid MNA-DPAs, while the dashed curves plot the case for an MNA in vacuum. We find that the radiative decay rate of the emitter is enhanced to about $5500\gamma_{\text{int}}$ or equivalently $970\gamma_{\text{vac}}$ at a wavelength of 580 nm and for $n$=2.5, compared to the radiative decay rate of only $20\gamma_{\text{vac}}$ for the same MNA-emitter separation in the absence of the DPA (Fig. 1a). In the case of semiconductors with $n = 3.5$, we reach $\gamma_{\text{rad}} = 3200\ \gamma_{\text{vac}}$, which is 29100 times higher than its value of $\gamma_{\text{int}} = 0.11\gamma_{\text{vac}}$ in the absence of the MNA. Furthermore, Fig. 2b reveals that a simple metallo-dielectric hybrid antenna provides much higher $\eta$ than a MNA alone. We note that the enhancement of $\gamma_{\text{rad}}$ strongly depends on $n$, whereby the inset of Fig. 2a shows the insensitivity of $\gamma_{\text{int}}$ to wavelength variations. We remark in passing that the concept presented here also applies to nanoscopic emitters such as diamond nanocrystals placed on dielectric surfaces [23, 24].

We now present physical arguments for the superior performance of the metallo-dielectric antenna and exploit them to improve it even further. Let us first consider the electric polarization (dipole moment per unit volume) [25] in the MNA written as $\mathbf{P} = \varepsilon_0(\varepsilon_{\text{Au}} - 1)\mathbf{E}(\vec{r})$. This results in an induced dipole moment given by

$$\mathbf{u} = \int \mathbf{P} dv = \int \varepsilon_0(\varepsilon_{\text{Au}} - 1)\mathbf{E}(\vec{r})dv\ , \tag{1}$$

leading to the radiated power [25, 26] $P_{\text{rad}} = \frac{c_0^2 Z_0 k_0^4}{12\pi}|\mathbf{u}|^2 \delta$. Here $c_0$ and $Z_0$ stand for the speed of light and impedance in vacuum, respectively, and $\delta$ represents the LDOS normalized by its value in vacuum. Knowing the dependence of $\delta$ on the proximity of a dipole to an interface [17], we can assign a value to $\delta$ at the location of the MNA estimated as $\bar{z} = \left|\frac{\int \mathbf{P} \cdot z dz}{\int \mathbf{P} \cdot dz}\right|$. For the arrangement shown in Fig. 1c, one finds $\bar{z} = 39$ nm, leading to $\delta = 2.4$ for $n$=2.5 at $\lambda = 580$ nm and $\delta = 2.7$ for $n$=3.5 at $\lambda = 664$ nm.

Next, we examine $\mathbf{u}$ in Eq. (1). Noting that only $E_z$ contributes to the induced dipole moment (due to the rotation symmetry of our system), we express it in cylindrical coordinates $(\rho, z)$ as $E_z = q(\rho, z)E_0$, where $E_0$ is the maximum electric field amplitude in the MNA. We then define an effective polarizability $\alpha_e = (\varepsilon_{Au} - 1) \iint q(z,\rho) 2\pi\rho dz d\rho$ according to $\boldsymbol{u} = \varepsilon_0 \alpha_e E_0$. Figure 3a displays the rigorous BOR-FDTD computation of $|\int q(z,\rho) \rho d\rho|$ as a function of $z$. Compared to the case of no DPA ($n$=1), we find that the polarization drops more slowly within the MNA, inducing a greater $\alpha_e$ and thus more efficient radiation. Also, as seen from the intensity plots in Fig. 1, the MNA-DPA hybrid structures offer larger electric fields inside the metal and correspondingly much larger $\boldsymbol{u}$. Figure 3b depicts the wavelength dependence of the enhancement factor $\phi = u/u_0$ for the induced dipole, where $\boldsymbol{u_0}$ is the dipole moment in the absence of a DPA. We see that $\phi$ has a resonance feature that shifts to longer wavelengths as the refractive index of the DPA increases. Noting that $P_{\text{rad}}$ is enhanced by $\phi^2 \delta$, one readily understands the origin of the resonances in Fig. 2a. For example, for the arrangement of Fig. 1c, the estimated value of $P_{\text{rad}}$ is enhanced (compared to the case without DPA) by 42 times for $n$=2.5 at $\lambda = 580$ nm and 213 times for $n$=3.5 at $\lambda = 664$ nm. These estimates are in good agreement with the enhancement factors of 49 and 239 calculated by rigorous BOR-FDTD. We remark that in our previous experiments [10, 27] we could not observe such large enhancements of spontaneous emission because of the limited instrumental time resolution. Moreover, in those efforts the gold nanoparticle was not brought to contact with the dielectric film [24].

The above discussion shows that the presence of a DPA increases both the dipole moment induced in a MNA and the LDOS at its position, resulting in stronger radiation. Another great advantage of the hybrid MNA-DPA antenna is the less prohibitive behavior of $\eta$. For instance, we obtain $\eta = 46\%$ for a DPA with $n = 3.5$ at $\lambda = 664$ nm as compared to $\eta = 16\%$ for an MNA alone. The unavoidable accompaniment of the nonradiative decay of the excited state [14, 28] is, indeed, a major challenge in the enhancement of spontaneous emission with plasmonic nanoantennas. Let us remember that $\eta$ can be calculated according to $\eta = \frac{P_{\text{rad}}}{P_{\text{rad}} + P_{\text{nr}}}$, where $P_{\text{nr}} = \frac{1}{2} Re\{\int \boldsymbol{E} \cdot \boldsymbol{J}^* dv\}$ computes the power dissipated in the MNA based on the induced current density $\boldsymbol{J}(r) = -i\omega\varepsilon_0(\varepsilon_{\text{Au}} - 1)\boldsymbol{E}(\vec{r})$ inside it. The quantum efficiency $\eta$ can be then written in the form $\eta = \frac{1}{1+C/(\delta \cdot f)}$ with $f = \frac{|\iint q(z,\rho)\rho dz d\rho|^2}{\iint |q(z,\rho)|^2 \rho dz d\rho}$ and $C = \frac{6\pi\omega \text{Im}[\varepsilon_{Au}]}{\varepsilon_0 c_0^2 Z_0 k_0^4 |\varepsilon_{Au}-1|^2}$. Since the latter depends only on the wavelength and the dielectric function of gold, the role of the geometry and the spatial distribution of the field enter via $f$. For the examples shown in Fig. 3a, one obtains $\frac{f_{2.5}}{f_{1.0}} = 2.6$ at $\lambda = 580$ nm and $\frac{f_{3.5}}{f_{1.0}} = 3.0$ at $\lambda = 664$ nm, confirming that a more homogeneous distribution $q(z,\rho)$ offers a larger $f$ and, therefore, higher quantum efficiency.

We have shown that a simple metallo-dielectric antenna consisting of a spherical MNA and a dielectric interface as DPA can substantially improve the performance of plasmonic nanoantennas in the wavelength range where metals absorb strongly. What we have learned can be summarized in the following guidelines: *i*) maximize the field driving the MNA; *ii*) distribute the field in the latter as uniformly as possible to obtain a large dipolar polarizability; *iii*) place the MNA in a location where the LDOS is high so that the dipole induced in the MNA radiates efficiently. Equipped with these design rules, we can now achieve larger effects by optimizing the MNA and DPA constituents.

In a recent work, we showed that single nanocones act as very efficient MNAs in vacuum [29], yielding higher enhancements than nanospheres, rods, or ellipsoids. Figures 4a and 4b show $\gamma_{\text{rad}}$ and $\eta$ as a function of the wavelength for a truncated conical gold nanoantenna in contact with substrates of $n=2.5$ and $n=3.5$, respectively. We find that this simple arrangement yields enhancements of 5360 and 7830 times over $\gamma_{\text{vac}}$ at resonance, corresponding to enhancements of 31000 and 67500 over $\gamma_{\text{int}}$, i.e. without MNA. As before, we manage to maintain $\eta>70\%$, which is more than twice the quantum efficiency of an emitter coupled to a bare MNA. We note that as seen in the inset of Fig. 4a, we considered the MNA to have a flattened end to approximate realistic laboratory conditions. The dip in the quantum efficiency curve for $n=3.5$ (Fig. 4b) is caused by this flattening, which leads to a localized multipolar resonance and lower radiation efficiency [24]. Our calculation showed that the maximum enhancement factor varied by ±20% for tip end curvatures ranging from 1 nm to 5 nm. The surface roughness at other parts of the cone should be less critical as long as the aspect ratio of the cone is not changed. Another sensitive parameter in the performance of plasmonic nano-antennas is its distance to the emitter, which influences the enhancement factor exponentially [24]. To this end, in the case of emitters such as large quantum dots their finite size would have to be taken into account.

To obtain even higher quantum efficiencies, one can increase the LDOS further at the position of the plasmonic nanoantenna to promote its radiation compared to dissipation. As discussed in the context of CQED [2, 18], this can take place in the gap between two dielectric media. For example, the emission of an atom at a wavelength of 700 nm is enhanced by about 8 times if it is placed at the center of a gap of 80 nm between two dielectric half spaces of $n$=3.5. We have, thus, evaluated the performance of a hybrid metallo-dielectric antenna geometry illustrated in the inset of Fig. 4c. The data in Fig. 4 show that $\eta$ can reach up to 85% for $n$=3.5 in a wide spectral range although in this case $\gamma_{\text{rad}}$ settles at a lower value of $3000\gamma_{\text{vac}}$. Aside from providing a larger $\eta$, this arrangement has the important practical advantage that the MNA can be fabricated on a dielectric slab such as a cantilever/tip combination and be positioned at will on a substrate [10].

We have presented design concepts for improving plasmonic nanoantennas by combining them with dielectric antenna structures, whereby we have discussed examples with experimentally realistic parameters such as antenna geometry and its distance to the emitter. Recent reports predict ten to a few-hundred fold enhancement of single-photon emission for emitters in dielectrics by using metallic cylindrical cavity structures [30-31] and patch antennas [32]. In our work, the resulting hybrid antennas open the door to an unprecedented paradigm, where the decay of the excited states of optical emitters can be sped up by tens of thousands for nanoscopic emitters in dielectrics close to the interface. For example, the radiative lifetime of semiconductor quantum dots can be shortened from a few ns to several 100 fs. This makes it possible to cycle an emitter much faster and receive more single photons per unit time, greatly facilitating detection and spectroscopy of single solid-state emitters such as ions. The radiation pattern of the metallo-dielectric antenna is similar to that of a vertical dipole antenna above a high-index substrate. About 82% of the emitted photons funnel to the substrate. Furthermore, such bright emission combined with near-unity collection schemes [21] brings about the prospect of triggered single-photon sources at the μW power level for various applications [33].

Sub-picosecond radiative processes in the solid state are also interesting because they become comparable with or faster than many vibrational and phononic interactions, leading to new regimes of photophysical dynamics. For instance, coherent coupling of multichromophore systems [34], efficient emission from quenched, blinking, and dephased systems, as well as the strong coupling between an emitter and plasmons [35] become possible. Another very exciting implication of our work is in surface-enhanced Raman spectroscopy (SERS) at the single-molecule level [36], where the metallo-dielectric hybrid antenna could yield electromagnetic enhancement factors of the order of $10^8$-$10^{10}$ when considering that reciprocity translates the huge enhancements of spontaneous emission to similar electric field intensity amplifications in the excitation channel. The great advantage of our scheme is that it allows non-contact Raman spectroscopy even on single molecules that are embedded under a surface. In closing, we anticipate that further optimization of the MNA (e.g. using silver) as well as the structure of the DPA (e.g. using layered antennas) can give rise to radiative lifetimes below 100 fs [24].

**Figures and captions**

Figure 1

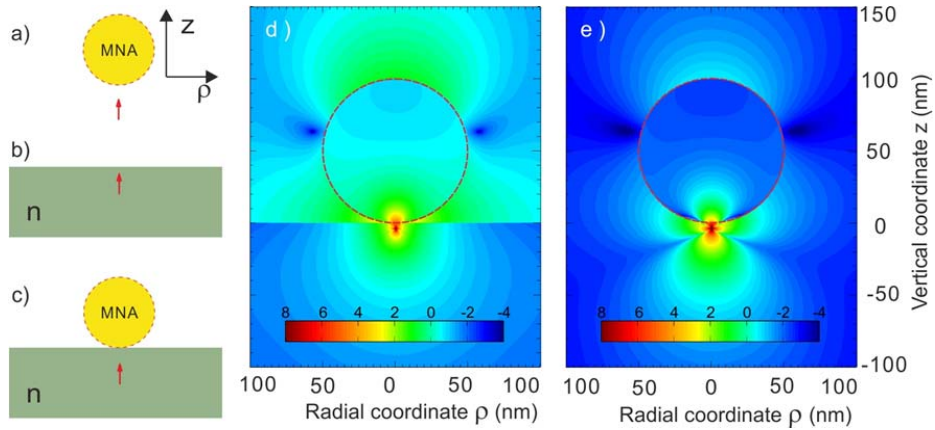

**Figure 1** a) Metallic nanoantenna (MNA) and emitter in vacuum; b) Emitter embedded in a dielectric planar antenna (DPA) with refractive index n; c) Emitter coupled to a MNA-DPA hybrid antenna; d, e) Intensity profiles of the *z*-component of the electric field generated by an oscillating electric dipole oriented along *z* coupled to MNA-DPA and MNA, respectively. The MNA-emitter distance is fixed to 4 nm in both cases, the refractive index of the substrate is 2.5, and the emission wavelength is set at 580 nm. The intensity plots follow logarithmic scales.

Figure 2

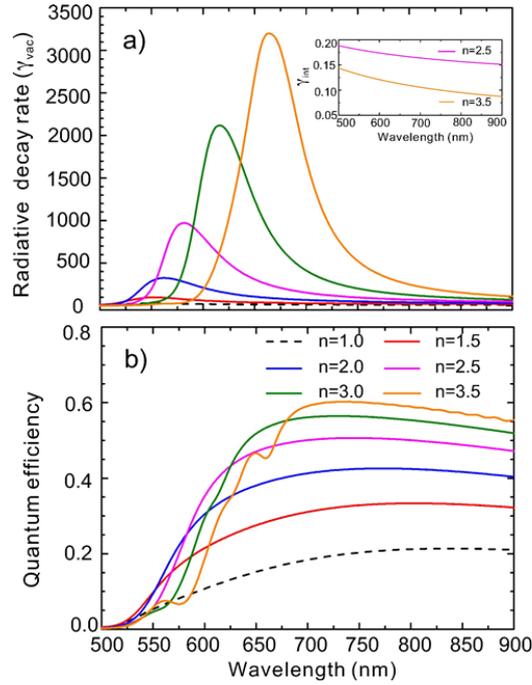

**Figure 2** The radiative decay rate normalized to $\gamma_{vac}$ (a) and quantum efficiency (b) as a function of the emission wavelength. The emitter is embedded in dielectric substrates of different *n* and placed at *d* = 4nm from a spherical MNA (see Fig. 1c). The dashed lines plot the results for a MNA alone. The local field effect on the decay rate is neglected [15, 16], and the initial value of $\eta$ is assumed to be 100% for the naked emitter. The inset in (a) shows the spectra of $\gamma_{int}$ for two different *n*. The small modulations on the curve for *n*=3.5 in (b) is caused by a residual effect of the finite grid size, which excites localized multipolar resonances [24].

Figure 3

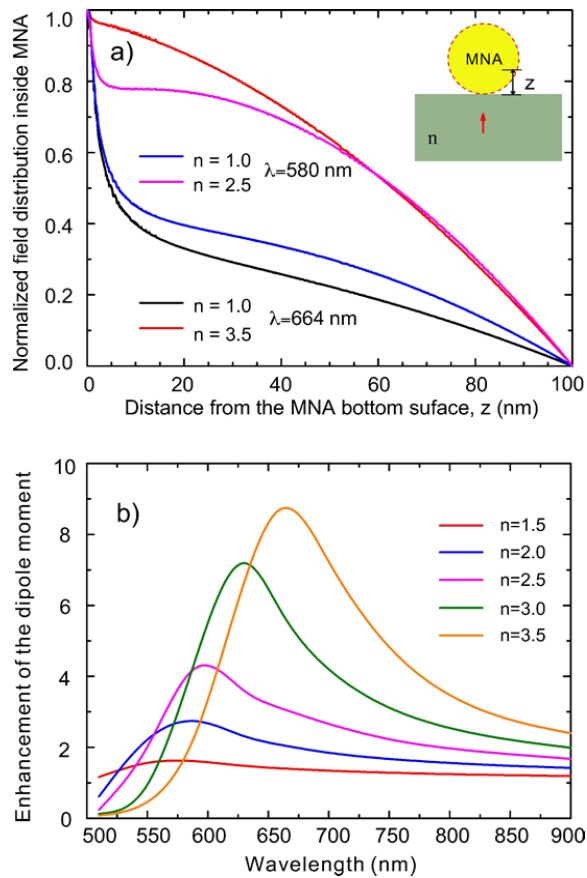

**Figure 3** (a) Distribution of the field amplitude (the quantity $|\int q(z,\rho)\,\rho d\rho|$, see text for details) induced inside a spherical MNA normalized to its maximum for DPAs of different dielectric indices at two wavelengths. It is clear that the field is more uniformly distributed in the presence of a dielectric interface. (b) Spectra of the enhancement factor ($u/u_0$) of the induced dipole moment in the MNA for various refractive indices of DPA.

Figure 4

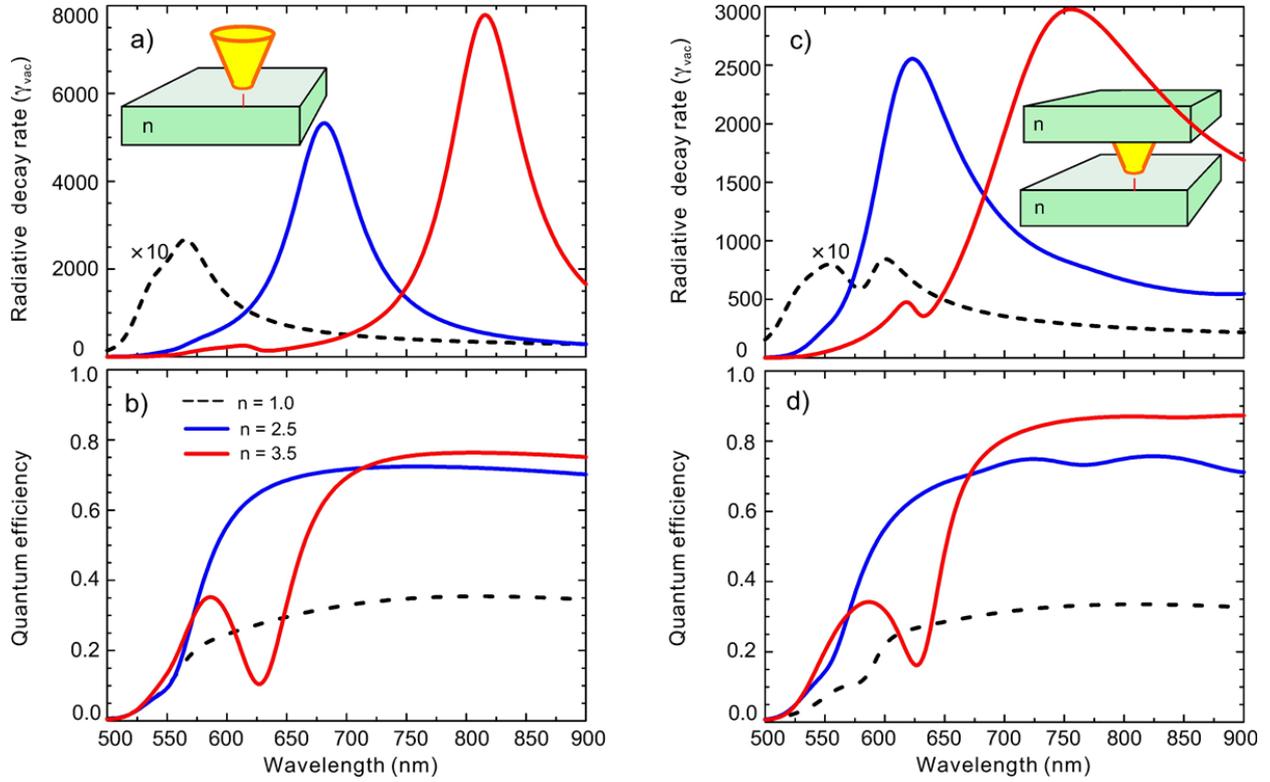

**Figure 4** Radiative decay rate in unit of $\gamma_{\text{vac}}$ (a, c) and quantum efficiency (b, d) as a function of wavelength for two types of MNA-DPA hybrid antennas. The gold nanocone in (a) is 80 nm long and starts with a base radius of 80 nm and finishes with a flattened tip end of 3 nm radius in contact with the DPA. The nanocone in (c) is the same but starts with a larger base radius of 120 nm. Both sides of the MNA are in contact with the dielectric substrates. The local field effect on the decay rates was neglected [15, 16].